\documentstyle[aps,twocolumn,epsfig]{revtex}

\begin{document}
\draft

\twocolumn[\hsize\textwidth\columnwidth\hsize\csname@twocolumnfalse\endcsname

\title{Quantum entanglement and Bell inequalities  in   Heisenberg spin chains}
\author{Xiaoguang Wang and  Paolo Zanardi}
\address{Institute for Scientific Interchange (ISI) Foundation,\\
 Viale Settimio Severo 65, I-10133 Torino, Italy}
\date{\today}
\maketitle

\begin{abstract}
We show that in one-dimensional 
isotropic Heisenberg model  two-qubit thermal entanglement and maximal violation of 
Bell inequalities 
are directly related with a   thermodynamical
state function, i.e., the internal energy.
Therefore they are completely determined by
the partition function, the central object of thermodynamics. 
For ferromagnetic ring we prove that  there is no thermal entanglement at any temperature. 
Explicit relations between the concurrence and the measure of maximal Bell inequality
violation are  given. 
\end{abstract}
\pacs{PACS numbers: 03.65.Ud, 75.10.Jm }
]

Over the past few years much effort has been put into studying the
entanglement of quantum systems both qualitatively and quantitatively.
Entangled states constitute indeed a valuable resource in quantum information
processing \cite{Bennett}. 
Entanglement in   systems of interacting spins  
 \cite
{Connor01,Meyer01,Arnesen01,Wang01,Three}
as well as in systems of indistinguishable particles \cite{Schli,Li,Paolo,Paolo1,You}
has been investigated. 
In particular entanglement in both the
ground state \cite{Connor01,Meyer01} and thermal state \cite
{Arnesen01,Wang01,Three} of a spin-1/2 Heisenberg spin chain have been  analyzed
in the literature. The  intriguing
issue of the relation between entanglement and
quantum phase transition \cite{QPT} have been  addressed  in a few quite recent
papers \cite{Osborne,Osterloh}.

{}From a conceptual perspective one can say that 
these  investigations  aim to provide a bridge 
between quantum
information theory and theoretical  condensed matter physics. 
This is done by considering 
 thermal equilibrium in a canonical ensemble.
In this situation the system state is given
by the Gibb's density operator $\rho _T=\exp
\left( -H/kT\right) /Z,$ where $Z=$tr$\left[\exp \left( -H/kT\right) \right] 
$ is the partition function, $H$ the system Hamiltonian, $k$ is Boltzmann's
constant which we henceforth will take equal to 1, and $T$ the temperature. As $%
\rho (T)$ represents a thermal state, the entanglement in the state is
called {\em thermal entanglement}\cite{Arnesen01}. It is important to stress
that,
although the central object of  statistical physics, 
the partition function, is  determined by the  eigenvalues of $H$ only,
thermal entanglement properties generally require in addition 
the knowledge of the 
energy eigenstates. 

On the other hand the violation of Bell inequalities \cite{Bell} was
considered as a means of determining whether there is entanglement. In 1989
Werner \cite{Werner89} demonstrated that that there exists states which are
entangled but do not violate any Bell--type inequality, i.e., not all
entangled states violate a Bell inequality. Further studies \cite
{Popescu,Gisin} showed that the maximal violation of a Bell inequality does
not behave monotonously under local operation and classical communications.
So Bell violations  can only be considered
an entanglement witness \cite{Terhal}. More recently 
the  relation between Bell inequalities and the usefulness for quantum key
distribution and quantum secret sharing have been clarified \cite{QKD}. 
Moreover  it has been  showed that
even multipartite bound entangled states can violate Bell inequalities \cite
{Dur}.

In this paper we shall study the  Heisenberg qubit chains by exploring further
connections between entanglement and other relevant  physical
quantities. We will study the   relations between   the concurrence, an entanglement
measure for two qubits, with thermodynamic potentials  such as the internal energy
 $U$ and magnetization $M.$ The latter quantities  are defined as 
\begin{equation}
U=-\frac 1Z\frac{\partial Z}{\partial \beta },\text{ }M=-\frac 1{Z\beta }%
\frac{\partial Z}{\partial B},
\end{equation} 
where $\beta=1/T$ and $B$ is a magnetic field.
We shall establish   a remarkably simple  relation between the concurrence 
for the
thermal state and the internal energy. A direct relation between the maximal
violation of the Bell inequality and the internal energy
will be given as well.

We first briefly review the definition of concurrence \cite{Con}. Let $\rho
_{12}$ be the density matrix of a pair of qubits $1$ and $2.$ The density
matrix can be either pure or mixed. The concurrence corresponding to the
density matrix is defined as 
\begin{equation}
C=\max \left\{ \lambda _1-\lambda _2-\lambda _3-\lambda _4,0\right\} ,
\label{eq:c1}
\end{equation}
where the quantities $\lambda _1\ge \lambda _2\ge \lambda _3\ge $ $\lambda
_4 $ are the square roots of the eigenvalues of the operator 
\begin{equation}
\varrho _{12}=\rho _{12}(\sigma _y\otimes \sigma _y)\rho _{12}^{*}(\sigma
_y\otimes \sigma _y).  \label{eq:c2}
\end{equation}
The nonzero concurrence implies that the qubits 1 and 2 are entangled. The
concurrence $C_{12}=0$ corresponds to an unentangled state and $C_{12}=1$
corresponds to a maximally entangled state.

{\em The Heisenberg model and  pairwise entanglement}. We consider a
physical model of a chain of $N$ qubits, namely, a chain of spin-$\frac 12$
particles in which neighboring particles interact via the anisotropic
Heisenberg Hamiltonian $H=H(\Delta ,B),$ with a magnetic field 
\begin{eqnarray}
H =J\sum_{i=1}^N\left( \vec{\sigma}_i\cdot \vec{\sigma}%
_{i+1}+(\Delta -1)\sigma _{iz}\sigma _{i+1z}\right)  
+B\sum_{i=1}^N\sigma _{iz},
\end{eqnarray}
where $\vec{\sigma}_i=(\sigma _{ix},\sigma _{iy},\sigma _{iz})$ is the
vector of Pauli matrices and $J$ is the exchange constants. The positive and
negative $J$ correspond to the antiferromagnetic (AFM) and ferromagnetic
(FM) case, respectively. 

In order to enact translational invariance for finite $N$
 we assume  periodic boundary conditions, i.e., $N+1\equiv 1,$ 
turning the chain into a ring.
Therefore $H$  becomes invariant under cyclic shifts generated by the 
right shift operator $T.$ The latter 
being  defined by its action on the product basis, $%
T\,|m_1,...,m_{N-1},m_N\rangle =|m_N,...,m_1,m_{N-1}\rangle ,$ where $m_i=0$
(1) represents the state of spin up (down). 
Another fact is that that $[H,S_z]=0,$ which
guarantees that reduced density matrix $\rho _{12}$ of two nearest-neighbor
qubits, say qubit 1 and 2, for the thermal state has the form \cite{Connor01}
\begin{equation}
\rho _{12}=\left( 
\begin{array}{llll}
u^+ &  &  &  \\ 
& w_1 & z^{*} &  \\ 
& z & w_2 &  \\ 
&  &  & u^-
\end{array}
\right)   \label{eq:rho12}
\end{equation}
in the standard basis $\{|00\rangle ,|01\rangle ,|10\rangle ,|11\rangle \}.$
Here $S_\alpha =\sum_{i=1}^N\sigma _\alpha /2$ $(\alpha =x,y,z)$ are the collective spin operators. The reduced density matrix is directly related to various correlation
functions $G_{\alpha \beta }=\langle \sigma _{1\alpha }\sigma _{2\beta
}\rangle =$tr$(\sigma _{1\alpha }\sigma _{2\beta }\rho _T)$. Then matrix
elements can be written in terms of the correlation functions and the
magnetization per site $\bar{M}=M/N$ as 
\begin{eqnarray}
u^{\pm} &=&\frac 14(1\pm2\bar{M}+G_{zz}),  \nonumber \\
z &=&\frac 14(G_{xx}+G_{yy}+iG_{xy}-iG_{yx}).  \label{eq:uvz}
\end{eqnarray}
In the deriving of above equation, we have used the translation invariance
of the Hamiltonian. From the above equation we find the following relations
which will be useful  later 
$u^+-u^- =\bar{M},  
u^++u^- =\frac 12\left( 1+G_{zz}\right) ,  
\mathop{\rm Re}(z) =\frac 14(G_{xx}+G_{yy}).  
$
All the information needed is contained in the reduced density matrix, from
which the concurrence quantifying the entanglement is readily obtained as%
\cite{Connor01} 
\begin{equation}
C=2\max [0,|z|-\sqrt{u^+u^-})],
\end{equation}
which can be expressed in terms of the correlation functions and the
magnetization as seen from Eq.(\ref{eq:uvz}).

{\em Isotropic Heisenberg model.} Now we consider the  Heisenberg $XXX$
model described by the  Hamiltonian $H(1,0)$. We first notice that $H$
admits  a continuous  $SU(2)$ group of symmetries. 
This can be easily checked 
in that $H$ commutes with the Lie-algebra generators
$S_\alpha,$
In particular the isotropic Hamiltonian also commutes
with the operators
\begin{equation}
Q_\alpha =\sigma _\alpha ^{\otimes N}\,(\alpha=x,y,z)
\end{equation}
that  generate $Z_2$ sub-groups of $SU(2).$ 
The  rich rotational symmetry structure of the $XXX$ model along with
 translational invariance   will play a key role in  our 
 study of the ring entanglement. 

Form the  $Z_2$ symmetry immediately follows  

{\em Proposition 1.} For any temperature the magnetization $M$ is vanishing
and the nondiagonal element $z$\ is real.

{\em Proof:} 
By definition one has 
$
M =N\langle \sigma _{1z}\rangle =NZ^{-1}\text{tr(}\sigma _{1z}e^{-\beta H}) 
=NZ^{-1}\text{tr(}\sigma _x^{\otimes N}\sigma _{1z}\sigma _x^{\otimes
N}e^{-\beta H})  
=-NZ^{-1}\langle \sigma _{1z}\rangle =-M.
$
Moreover
$
z =\langle \sigma _{1+}\sigma _{2-}\rangle =Z^{-1}\text{tr(}\sigma _{1+}\sigma
_{2-}e^{-\beta H})  
=Z^{-1}\text{tr(}\sigma _x^{\otimes N}\sigma _{1+}\sigma _{2-}\sigma _x^{\otimes
N}e^{-\beta H})  
=\langle \sigma _{2+}\sigma _{1-}\rangle =z^{*}.
$
$\hfill\Box$

In passing we notice that the reality of the correlation function 
$\langle\sigma_i^+ \sigma_j^-\rangle$ can be proven in a more general context
 possibly with  $Z_2$-symmetry broken. Indeed it is easy to check
that it is sufficient that there exist
a symmetry $S$ of the Hamiltonian swapping the $j$-th with the $i$-th
lattice site  i.e., $S\,\sigma_j^+\,S= \sigma_i^+.$
This is the case, for example, for the $XXX$ on a ring with  $S$
given by  a geometrical symmetry of the lattice, e.g., a reflection.

In view of  Prop. 1   the concurrence has the following simplified
form,
$C=2\max [0,|z|-u^+].$ Furthermore, from 
the relations for $u$, $v$, and $z$ established above  we 
find that the concurrence is
completely determined by the correlation functions $G_{\alpha \alpha
}(\alpha =x,y,z),$ i.e., 
\begin{equation}
C=\frac 12\max [0,|G_{xx}+G_{yy}|-G_{zz}-1)].  \label{eq:ccc}
\end{equation}
This result is  due to the $Z_2$ symmetry generated by $\sigma_x^{\otimes\,N}.$
By using the global $SU(2)$ 
 symmetry and the translation invariance it is starightforward to
check that $G_{xx}=G_{yy}=G_{zz}$ and $G_{xx}=U/(3JN)$. Then we get 

{\em Proposition 2}
 For the isotropic Heiserberg model the
concurrence of nearest--neighbor qubits is directly obtained from the
internal energy, i.e., 
\begin{equation}
C=\frac 16\max [0,2|\bar{U}/J|-\bar{U}/J-3)].  \label{eq:th1}
\end{equation}

In Ref. \cite{Connor01}  it has been shown that in the Heiserberg even-site
antiferromagnetic ($J=1$) $XXX$ model \cite{Connor01} there exists a direct
relation between the pairwise entanglement and the ground state (GS) energy $%
E_{GS}$: 
\begin{equation}
C=\frac 12\max \left( 0,-{E_{GS}}/N-1\right) .  \label{eq:w}
\end{equation}
The key step for the proof of this relation is to prove that the nondiagonal
element $z<0$ for the ground state. Note that this relation is obtained for
the case of antiferromagnetic and even-number qubits. One can  guess that the
nondiagonal element $z$ is also negative for the thermal state of the
antiferromagnetic model. In fact we can prove a more general result given by

{\em Proposition 3 } The internal energy is always
negative and the off--diagonal element $z$ is negative (positive) for the
antiferromagnetic (ferromagnetic) rings.

{\em Proof:}
 First, from the tracelessness of the Pauli operators, it is immediate  to check  that in the limit of 
$T\rightarrow \infty $ one has  $U=0.$ The  internal energy is known to be a non-decreasing
function of the temperature 
($\partial U/\partial T={(\Delta H)^2}/{T^2}\geq 0$)
 $U$ is
negative. Since we know that $z=U/(6JN),$ we conclude that $z<0$ for the
antiferromagnetic case and $z>0$ for the ferromagnetic case $\hfill\Box$

The combination of the Prop. 2 and 3 gives rise to 

{\em Theorem 1} The concurrence of the nearest-neighbor qubits in the
Heisenberg model is given by 
\begin{equation}
C=\left\{ 
\begin{array}{ll}
\frac 12\max [0,-\frac U{JN}-1] & ~~\text{for AFM,} \\ 
\frac 12\max [0,\frac U{3JN}-1] & ~\text{~for FM.}
\end{array}
\right.   \label{eq:cxxx}
\end{equation}

In the limit $T\rightarrow 0$ in the antiferromagnetic even-site model, the
thermal state become the nondegenerate ground state. As we expected Eq.(\ref
{eq:cxxx}) reduces to Eq.(\ref{eq:w}) in this limit. Theorem 1 establishes  a
direct relation between the concurrence and a  macroscopic thermodynamical
function, the internal energy. Then the entanglement is uniquely determined
by the partition function of the system. Notice  that  Theorem 1 has no
restriction to $J>0$ and $N$  even (like in Ref \cite{Connor01}).
Eqs. (\ref{eq:cxxx}) clearly show that in AFM (FM) rings the concurrence
is a not increasing (decreasing) function of the temperature.

{}From the obvious fact 
\begin{eqnarray}
& &|\langle \sigma_z\otimes\sigma_z\rangle|=Z^{-1} |\sum_i e^{-\beta E_i}
\langle i |\sigma_z\otimes\sigma_z|i\rangle|\le  
\\ \nonumber
& &Z^{-1}
 \sum_i e^{-\beta E_i} 
\|\sigma_z\otimes\sigma_z|i\rangle\|\le Z^{-1} \sum_i e^{-\beta E_i}=1,
\end{eqnarray}
one has   that 
$-1\leq  U/{3JN}=G_{zz}\leq 1.$
Therefore we obtain a general result which is applicable for both even and
odd number of qubits.

{\em Corollary 1} At any temperature there is no thermal entanglement 
between two qubits in the
FM $XXX$ Heisenberg rings.

Now we discuss the case of AFM. When the temperature increases the internal
energy will increase but the concurrence decrease. When the internal energy
arrives at $-NJ$ the concurrence becomes zero. 

The temperature $T_c$  at which
the concurrence vanishes is called the threshold temperature. 
In a short summary we have

{\em Corollary 2} At temperatures lower than $T_c$ 
there exists thermal entanglement between two qubits
 in the AFM $XXX$  Heisenberg rings. 

$T_c$  is determined by the
equation 
$u(N):= U(T_c)/(-NJ)=1.$
{}From the numerical evidences of Refs. \cite{Connor01} and \cite{Odd}
we conjecture that $u(N)$ is a non-increasing (non-decreasing) function of $N$
for $N$ even (odd). Therefore since it has been estimated $u(\infty)>1$ \cite{Connor01}
and $u(3)<1, u(5)>1$ \cite{Odd}  one  obtains that {\em the ground state of the AFM Heisenberg
is always entangled except for the  case $N=3$.} 

Now we give the examples of 2 and 3 qubit for the illustration of
our general results. For two-qubit model the partition function is given by $%
Z=3e^{-2\beta J}+e^{6\beta J}$, then the internal energy follows $%
6J(e^{-2\beta J}-e^{6\beta J})/Z$. From Eq.(\ref{eq:cxxx}) the concurrence
is found to be zero for the FM rings and $C=\max \{0,[\exp (8\beta
J)-3]/\,[\exp (8\beta J)+3]\}$ for the AFM \cite{Arnesen01}.   The threshold temperature is then determined by $\exp (8\beta J)=3$ and it is given by $T_c=8J/\ln 3.$
For three-qubit model the partition function is given by $8\cosh (3\beta J)$%
, and then the internal energy is $-3J\tanh (3\beta J),$ from which and Eq.(%
\ref{eq:cxxx}) we can find that the concurrence is zero for the FM rings as
we expected and $C=\max \{0,\tanh (3\beta J)-1\}/2.$ Hence we recover the
result that there is no pairwise thermal entanglement for the three-qubit
Heisenberg ring \cite{Three}. Next we discuss Bell inequality.

{\em Bell inequality.} The most commonly discussed Bell--inequality is the
CHSH inequality\cite{Bell,CHSH}. The CHSH operator ($\vec{a},\vec{a^{\prime }%
},\vec{b},\vec{b^{\prime }}$ are unit vectors) reads 
\begin{equation}
\hat{B}=\vec{a}\cdot \vec{\sigma}\otimes (\vec{b}+\vec{b^{\prime }})\cdot 
\vec{\sigma}+\vec{a^{\prime }}\cdot \vec{\sigma}\otimes (\vec{b}-\vec{%
b^{\prime }})\cdot \vec{\sigma}.
\end{equation}
In the above notation, the Bell inequality reads 
\begin{equation}
\left| \langle \hat{B}\rangle \right| =\left| \text{tr}(\rho \hat{B})\right|
\leq 2,
\end{equation}
where $\rho $ is an arbitrary two--qubit state. The maximal amount of Bell
violation of a state $\rho $ is given by \cite{Horo}

\begin{equation}
{\cal B}=2\sqrt{u+\tilde{u}},
\end{equation}
where $u$ and $\tilde{u}$ are the two largest eigenvalues of $T_\rho T_\rho
^{\dagger }.$ The matrix $T_\rho $ is 
 determined completely by the correlation functions being
 a $3\times 3$ matrix whose elements are $%
(T_\rho )_{nm}=$tr$(\rho _{12}\sigma _n\otimes \sigma _m).$ Here $\sigma
_1\equiv \sigma _x,\sigma _2\equiv \sigma _y,$ and $\sigma _3\equiv \sigma
_z $ are the usual Pauli matrices. We call the quantity ${\cal B}$ the
maximal violation measure, which indicates the Bell violation when ${\cal B}
>2 $ and the maximal violation when ${\cal B}=2\sqrt{2}.$ 
The violation measure is a function of the correlation functions
and does not depend on the magnetization.

For the isotropic Heisenberg model the matrix $T_\rho $ is easily obtained
as diag$[G_{xx},G_{xx},G_{xx}]$ and the violation measure becomes 
\begin{equation}
{\cal B}=2\sqrt{2}|G_{xx}|=\left\{ 
\begin{array}{ll}
\frac{-2\sqrt{2}U}{3JN} & \text{For AFM}, \\ 
\frac{2\sqrt{2}U}{3JN} & ~~~~\text{For FM}.
\end{array}
\right.  \label{eq:bxxx}
\end{equation}
As we expected, the violation measure is completely determined by the
internal energy just as the concurrence.

{}From Eqs.(\ref{eq:cxxx}) and (\ref{eq:bxxx}) we arrive at an explicit
relation between the concurrence and the violation measure

\begin{equation}
C=\left\{ 
\begin{array}{ll}
\frac 12\max [0,\frac{3{\cal B}}{2\sqrt{2}}-1] & ~~~\text{For AFM}, \\ 
\frac 12\max [0,\frac{{\cal B}}{2\sqrt{2}}-1] & ~~~\text{For FM}.
\end{array}
\right.
\end{equation}
{}From the above relation we know that the concurrence is larger than zero
when the Bell inequality is violated. When $2\sqrt{2}/3<{\cal B}\leq 2$ for
the AFM rings the state is entangled, but the Bell inequality is not
violated. The result is general for arbitrary $N.$ The maximal value of $%
{\cal B}$ is $2\sqrt{2},$ therefore the concurrence is zero for the FM
rings. Next we consider the general model with anisotropy and magnetic
fields.

{\em General models including anisotropy and magnetic fields.}
For $\Delta\neq 1$ one obtains an anisotropic model which has no longer
a global SU(2)- symmetry. 
The corresponding Hamiltonian $H(\Delta,0)$ 
still  commutes with the $z$ component of the total spin. So
the concurrence is given by Eq.(\ref{eq:ccc}). For this model we have the
following relations $\bar{U}/J=G_{xx}+G_{yy}+\Delta G_{zz},$ where $\bar{U}=U/N$ is the internal energy per site. The combination
of the relation and Eq.(\ref{eq:ccc}) gives 
\begin{equation}
C=\frac 12\max [0,|\bar{U}/J-\Delta G_{zz}|-G_{zz}-1)].
\end{equation}
Then if we calculate the concurrence, we need to know the correlation
function $G_{zz}$ and the partition function. The partition function itself
is not  sufficient for determining the entanglement. 
In particular for $XY$ model, i.e., $\Delta=0,$  the concurrence reduces
to $C=\frac 12\max [0,|\bar{U}/J|-G_{zz}-1)].$ We still need to know the
correlation function $G_{zz}$\cite{lieb} to calculate the concurrence.

Now we consider the $XXX$ model with a magnetic field, i.e., the Hamiltonian 
$H(1,B).$ Now the magnetization is no long zero and hence $u^+ \neq u^-.$ For
this model we have the relation 
$\bar{U}/J =G_{xx}+G_{yy}+G_{zz}+B\bar{M}=4\mathop{\rm Re}(z)+2(u^++u^-)-1+B\bar{M}
$. Here we have used the relation $G_{xx}+G_{yy}+G_{zz}=4\mathop{\rm Re}(z)%
+2(u^++u^-)-1.$ Since $u^+-u^-=\bar{M},$ we can express $u$ and $v\,$ in terms of $z$ 
,$\bar{U}/J,$ and $M.$ Finally the concurrence is 
$C =2\max \{0,|z|-\frac 14[(\bar{U}/J-4\mathop{\rm Re}(z)+1-B\bar{M})^2 
-4\bar{M}^2]^{1/2}\}.  
$ To calculate the entanglement we need to know, beside the partition function,
the non-diagonal element $z.$

{\em Comments}. Our discussions above are applicable to more general
Heisenberg Hamiltonians such as

\begin{equation}
{\cal H}=\sum_{i\neq j}(J_{i,j}^x\sigma _{ix}\sigma _{jx}+J_{i,j}^y\sigma
_{iy}\sigma _{jy}+J_{i,j}^z\sigma _{iz}\sigma _{jz}),  \nonumber
\end{equation}
where $J_{i,j}^\alpha (\alpha =x,y,z)$ are arbitrary exchange constants. For
this model we still have the $Z_2$ symmetry [$H,\sigma _x^{\otimes N}]=0.$
Therefore the concurrence for the two qubits $i$ and $j$ of the thermal state is given,
with an extra $i,j$ dependence, by Eq.(\ref{eq:ccc}).
 If $J_{i,j}^x=J_{i,j}^y=J_{i,j}^z,$ then it is easy to
check that the Hamiltonian ${\cal H}$ has a  global $SU(2)$-symmetry.
Therefore the concurrence becomes $C=\frac 12\max [0,2|G_{zz}|-G_{zz}-1)],$
which is determined solely by the correlation function $G_{zz}.$

{\em Conclusions. } In this paper we have  discussed  thermal 
entanglement and Bell inequality violation in the  Heisenberg qubit rings. 
In the isotropic case we found
that
there exists a simple  relation between the pairwise entanglement for 
the thermal Gibb's state  and the internal
energy. This  result is noteworthy and somewhat surprising
in that it allows to directly  
connect entanglement properties  with  a thermodynamical potential
and thus eventually with the partition function.
In particular  one  can conclude that, at any temperature (above the threshold
temperature),
no pairwise thermal entanglement exists 
in the
FM (AFM) rings.

 We  also determined a  simple relation between the measure of maximal
violation of Bell inequality and the internal energy. This in turn
allows to explicitly show the  relation between the
concurrence and the violation measure.
The key ingredient in our derivations has been the vast symmetry group
of the  $XXX$ model. The study of the relation between thermal
entanglement and thermodynamical
quantities for spin models  with a poorer symmetry structure
is, we believe,  an intriguing challenge for future investigations.

The authors thanks for the helpful discussions with Dr. Irene D'Amico and
Prof. V.E.Korepin, H. Fu and A. I. Solomon. This work has been supported by the 
European Community through grant IST-1999-10596 (Q-ACTA).

\end{document}